# Beat Phenomena in Metal Nanowires, and their Implications for Resonance-Based Elastic Property Measurements

*Haifei Zhan[1], Yuantong Gu*[1] and Harold S. Park[2]*

[1]*School of Chemistry, Physics and Mechanical Engineering, Queensland University of Technology, Brisbane 4001, Australia*

[2]*Department of Mechanical Engineering, Boston University, Boston, Massachusetts 02215, USA*

**\*Corresponding Author:** Dr. Yuantong Gu

**Mailing Address:** School of Chemistry, Physics and Mechanical Engineering,

Queensland University of Technology,

GPO Box 2434, Brisbane, QLD 4001, Australia

**Telephones:** +61-7-31381009

**Fax:** +61-7-31381469

**E-mail:** yuantong.gu@qut.edu.au

**Abstract.** The elastic properties of 1D nanostructures such as nanowires are often measured experimentally through actuation of the nanowire at its resonance frequency, and then relating the resonance frequency to the elastic stiffness using elementary beam theory. In the present work, we utilize large scale molecular dynamics simulations to report a novel beat phenomenon in [110] oriented Ag nanowires. The beat phenomenon is found to arise from the asymmetry of the lattice spacing in the orthogonal elementary directions of the [110] nanowire, i.e. the [$\bar{1}$10] and [001] directions, which results in two different principal moments of inertia. Because of this, actuations imposed along any other direction are found to decompose into two orthogonal vibrational components based on the actuation angle relative to these two elementary directions, with this phenomenon being generalizable to <110> FCC nanowires of different materials (Cu, Au, Ni, Pd and Pt). The beat phenomenon is explained using a discrete moment of inertia model based on the hard sphere assumption, the model is utilized to show that surface effects enhance the beat phenomenon, while the effect is reduced with increasing nanowires cross-sectional size or aspect ratio. Most importantly, due to the existence of the beat phenomena, we demonstrate that in resonance experiments only a single frequency component is expected to be observed, particularly when the damping ratio is relatively large or very small. Furthermore, for a large range of actuation angles, the lower frequency is more likely to be detected than the higher one, which implies that experimental predictions of Young's modulus obtained from resonance may in fact be under predictions. The present study therefore has significant implications for experimental interpretations of Young's modulus as obtained via resonance testing.

**Introduction**. Due to their intriguing mechanical, electrical, optical and thermal properties, nanowires (NWs) have been widely applied as active components of nanoelectromechanical systems (NEMS)[1], such as high frequency resonators[2], field effect transistors (FETs)[3], nano switches[4], and other devices[5-7]. NWs are commonly utilized as a vibrating beam in NEMS, where they are most useful when vibrating continuously at or near their resonant frequency. Very minute changes in the local environment, such as perturbations in forces, pressure or mass, can be detected by monitoring the corresponding changes in the resonance frequency of the NW[8]. This notion of nanowire-based NEMS has been successfully applied in atomic force microscopy (AFM) and various kinds of sensors and actuators[7, 9]. Therefore, it is of great scientific interest to characterize the vibrational properties of NWs.

There have been a number of experimental, theoretical and computational studies on the vibrational properties of NWs. On the experimental front, various researchers have studied the resonant frequencies of both metallic and semiconducting NWs[10-18]. On the theoretical and computational front, most researchers have utilized either surface-based extensions of continuum elasticity theory[19, 20] or multi-scale computational techniques[21-23] to examine surface stress effects on the resonant frequencies of NWs. Other researchers have utilized molecular dynamics (MD) simulations to study the vibrational properties of metal NWs[8, 24]. However, most of these studies have focused upon <100> NWs, which do not exhibit a beat phenomenon because the lattice spacing along orthogonal coordinate directions of the cross-section is the same. Furthermore, the continuum mechanics-based models do not capture the discrete lattice asymmetry that is required to induce the beat phenomena. Moreover, researchers have experimentally found that FCC metal NWs prefer to grow along the <110>, and not <100> axial direction, as previously reported for Au[25], Ni[26], Ag[27, 28], Cu[29] and Pd NWs[30]. Finally, most experimental studies of NW mechanical properties have been performed on non-<100> orientated NWs[18, 31-34]. Because of this, it is important to understand the mechanical properties for <110> oriented metal NWs as obtained through vibration, or resonance tests and simulations. We perform such a comprehensive investigation using large scale MD simulations in the present work.

**Computational details**. The vibrational study is based on a series of large-scale MD simulations that were carried out on [110] orientated Ag NWs with a circular cross-section. We utilized a circular, and not rectangular or other geometric cross-section because it enables us to demonstrate that the beat phenomenon arises from the asymmetry in lattice spacing in orthogonal directions along the cross-sectional (110) planes, and not due to any asymmetry of the cross-sectional geometry. For the circular Ag NWs, the length $L$ was chosen for all geometries to be 34.71 nm, while the diameter ranged from 4-8 nm. We modelled Ag using the embedded-atom-method (EAM) potential developed by Foiles et al.[35] This potential was fitted to a group of parameters, including cohesive energy, equilibrium lattice constant, bulk modulus, and others including a lattice constant $a$ which is chosen as 0.409 nm[36].

During each simulation, the NW was first created assuming bulk lattice positions, and then relaxed to a minimum energy state using the conjugate gradient algorithm, i.e. the length of the NW was allowed to decrease in response to the tensile surface stress. We then used the Nose-Hoover thermostat[37, 38] to equilibrate the NW at a constant temperature (NVT ensemble) for 200 picoseconds (psec) at a time step of 2 femtoseconds while holding the newly obtained length of the NW fixed. Finally, the NWs are actuated by applying a sinusoidal velocity excitation $v(z) = \lambda \sin(kz)$ along the $x$-axis, where $\lambda$ is the actuation amplitude, and $k$ equals $\pi / L$. As illustrated in Fig. 1, the two ends of the NW are fixed in all three directions to mimic a doubly clamped beam, and the two lateral directions along the $x$ and $y$-axes are chosen as $[\bar{1}1\bar{2}]$ and $[\bar{1}11]$ (this coordination group is referred as C0), respectively. No periodic boundary conditions were utilized at any point during the simulation process.

We also emphasize that the NW was modeled using an energy-conserving (NVE) ensemble during the free vibration process following the velocity actuation, and that the applied velocity field increased the total potential energy by less than 0.1% (corresponding to an initial transverse displacement less than 2% of the nanowire length), which ensures that the oscillations occur in the linear regime. The overall simulation methodology to study the oscillatory properties of the NWs is identical to that used previously for metal NWs[8, 23, 39]. All simulations were performed using the open-source LAMMPS code developed at Sandia National Laboratories[40].

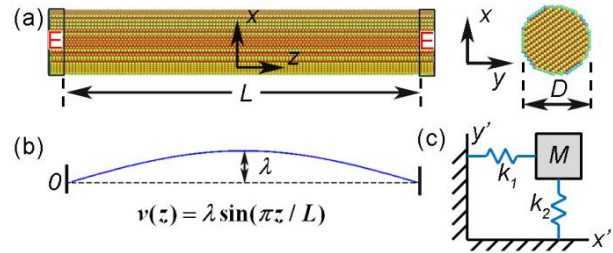

**Fig. 1.** (a) Schematic of a doubly clamped Ag NW with a circular cross-sectional area. The boundary regions 'E' are fixed in all directions, while the remainder of the NW is allowed to move freely. (b) The profile of the sinusoidal velocity actuation $v(z) = \lambda \sin(kz)$ imposed along the $x$ direction. (c) Simplified spring-mass model of the NW where $k_1$ and $k_2$ denote the effective elastic constants along the $x'$ and $y'$-axes, respectively.

**Results.** Figure 2a shows the time history of the external energy (*EE*) during the free vibration, where the *EE* is defined as the difference in the potential energy before and after the transverse velocity actuation is applied to the NW[8, 23]. As clearly demonstrated in Fig. 2a, the *EE* amplitude exhibits a periodic pulsation pattern, which is well-known from classical vibration theory[41] as arising due to a beat vibration or beat phenomena. Utilizing the fast Fourier transform (FFT)[42], the two frequency components that comprise the beat are identified in Fig. 2b as 22.07 and 25.12 GHz for the *EE* spectrum in Fig. 2a. This result indicates that the NW is under a combined vibration that is comprised of two orthogonal vibrational components whose frequencies are about 11.04 GHz and 12.56 GHz, respectively. Fig. 2c presents the atomic configurations of the NW at the time of 152 psec. As is seen, deflections along both the *x* and *y*-axes are observed. Therefore, we find that by imposing a single velocity actuation along the $[\bar{1}1\bar{2}]$ direction, the [110] orientated Ag NW is under a beat vibration state.

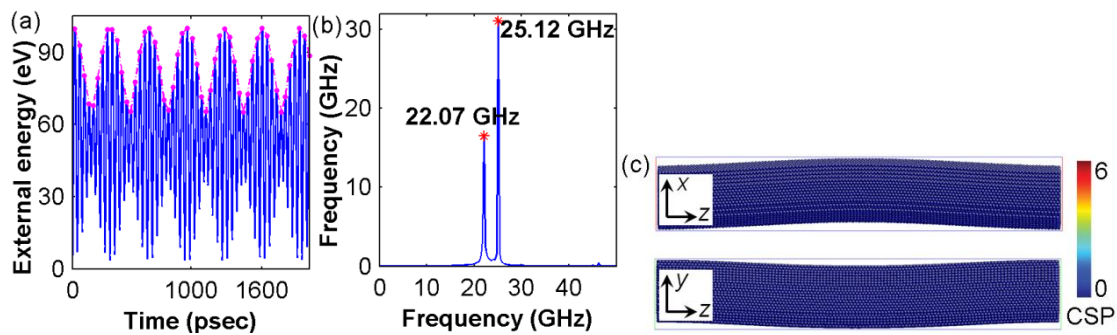

**Fig. 2.** (a). Time history of the external energy for a free vibration simulation at 10 K (time is truncated at 2000 psec). Circle markers are used as a guide to highlight the magnitude of the external

energy. (b) The frequency spectrum of the external energy using FFT analysis (truncated at a frequency of 50 GHz). (c) The atomic configurations of the NW at 152 psec. The upper figure shows the deflection along the *x*-axis, while the bottom figure illustrates the defection along the *y*-axis. Atoms are colored according to the centro-symmetry parameter (CSP)[43] between 0~6, which shows only the bulk atoms.

To further validate the existence of the beat phenomenon and also its generality, we performed many simulations of [110] Ag NWs considering different cross-sectional sizes, cross-sectional geometries (rhombic, triangular and pentagonal), actuation amplitudes, actuation along the *y*-axis, temperatures, FCC materials (Cu, Au, Ni, Pd and Pt), amounts of tensile and compressive pre-strain, methods of actuating the NW free vibration, i.e. using a cylindrical tip to indent the NW and drive the oscillation, and boundary conditions (i.e., fixed-free, or cantilevered). In all of these cases, as summarized in the Supporting materials, despite exhibiting different resonance frequencies, the same beat phenomenon was observed. Because of the apparent generality of the phenomena for [110] FCC metal NWs, we now proceed to explain its origin.

**Discussion.** We focus on explaining the mechanisms underlying the observed beat phenomenon, where a beat vibration is a combination of two simple harmonic vibrations with almost the same frequency. According to the vibration results in Fig. 2 and classical beam theory[41], the NW is vibrating via the first resonant mode in the transverse (*x* and *y*) directions and the displacement profile of the NW in each direction can be described by a sinusoidal function. For simplicity, we assume that the axial vibration is negligible and interpret the NW as a spring-mass system as illustrated in Fig. 1c. Considering these two orthogonal frequency components, two elastic constants are assumed along the $x'$ and $y'$-axes. Therefore, for a harmonic vibration, the displacement can be expressed as $U_i = A_i \sin(\omega_i t)$, where *i* equals 1 and 2 referring to the $x'$ and $y'$-axes, $\omega_i$ is the angular frequency of the NW and $A_i$ is the amplitude with no phase lag. Based on energy conservation arguments, i.e. that the maximum potential energy equals the maximum kinetic energy, the total external energy of the NW with mass *M* can be written as:

$$EE = \sum_{i=1,2} \frac{1}{2} M A_i^2 \omega_i^2 \sin^2(\omega_i t) \quad (1)$$

As demonstrated in Fig. 2b, the difference between the two frequencies components are relatively small, which means $\sin(\omega_1 - \omega_2)$ tends to zero. Therefore, equation (1) can be simplified as:

$$EE = A_{etot} - A_{etot} \cos[(\omega_1 - \omega_2)t]\cos[(\omega_1 + \omega_2)t] \quad (2)$$

where $A_{etot}$ is the amplitude of the total external energy, and $A_{etot} = (MA_1^2\omega_1^2 + MA_2^2\omega_2^2)/4$. Apparently, equation (2) describes a nearly harmonic vibration, with the amplitude oscillating under a low frequency, which is regarded as the beat vibration. We note that the beat vibration discussed here is generated by a single actuation. This is different from a continuous system, in which the beat vibration is usually generated by a harmonic actuation with a frequency close to the natural frequency of the system, i.e. the beat frequency equals $f_b = |\omega_1 - \omega_2|/(2\pi)$. Using the *EE* curve in Fig. 2a, the beat period is estimated to be 318 psec, which agrees well with the value of about 328 psec that is obtained from the frequency spectrum in Fig. 2b. This analysis demonstrates that the two natural frequencies of the NW observed in the MD simulations satisfy the fundamental requirements for a beat vibration.

To help resolve the issue as how two orthogonal vibrations can be excited in the [110] NWs by a single actuation, we note that due to the different atomic spacing along the *x* and *y*-directions for the cross-section of the [110] orientated Ag NWs shown in Fig. 3, we propose that the [110] orientated Ag NW possesses two orthogonal elementary orientations, and initial actuations imposed along any other direction will be decomposed into these two orthogonal actuations, which trigger two harmonic vibrations and lead to the beat phenomena. We have carried out several vibration tests to investigate this hypothesis. To do so, we specify four different orthogonal coordinate systems as illustrated in Fig. 3, which are denoted as C1 ([$\bar{1}$10] and [001]), C2 ([$\bar{5}$52] and [1$\bar{1}$5]), C3 ([$\bar{3}$32] and [1$\bar{1}$3]) and C4 ([$\bar{5}$56] and [3$\bar{3}$5]). Vibration tests were conducted by imposing a single velocity actuation to the NW at 10 K for each of these orthogonal coordinate systems C1-C4.

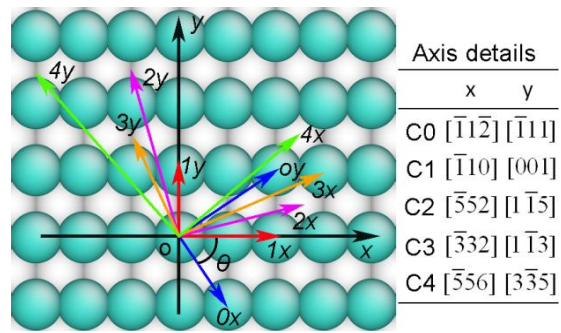

**Fig. 3.** (110) cross-sectional plane of the [110] Ag NW showing five groups of different orthogonal coordination systems as listed in the left table, including C0 ([$\bar{1}$1$\bar{2}$] and [$\bar{1}$11]), C1 ([$\bar{1}$10] and [001]), C2 ([$\bar{5}$52] and [1$\bar{1}$5]), C3 ([$\bar{3}$32] and [1$\bar{1}$3]) and C4 ([$\bar{5}$56] and [3$\bar{3}$5]). The angles between each direction with the [$\bar{1}$10] direction are denoted as the actuation angle $\theta$.

We find in conducting this series of tests that no obvious oscillation of the *EE* magnitude is observed for the C1 coordinate system, and that the beat phenomenon appears more and more apparent with an increase in the actuation angle $\theta$ (see Supporting materials for more details), where $\theta$ is the angle between the actuation direction and the [$\bar{1}$10] direction (see Fig. 3). These results signify that no beat phenomenon occurs for actuation along C1, while clear evidence of the beat phenomena is observed for actuation along C4, and verifies our hypothesis that the $[\bar{1}10]$ and $[001]$ directions are the two orthogonal elementary orientations, and initial actuations imposed along any other directions will be automatically decomposed based on these two elementary orientations. Further evidence of this conclusion can be found from comparisons of the natural frequency values extracted from the testing results in Fig. 4. As illustrated in Fig. 4, only one frequency value is identified for the $[\bar{1}10]$ and $[001]$ loading directions, whereas all other loading directions exhibit two frequency components. Importantly, these two frequency values are nearly identical to the values for the $[\bar{1}10]$ and $[001]$ actuation directions, which again demonstrates that these two directions are the two elementary orientations for $[110]$ orientated Ag NWs.

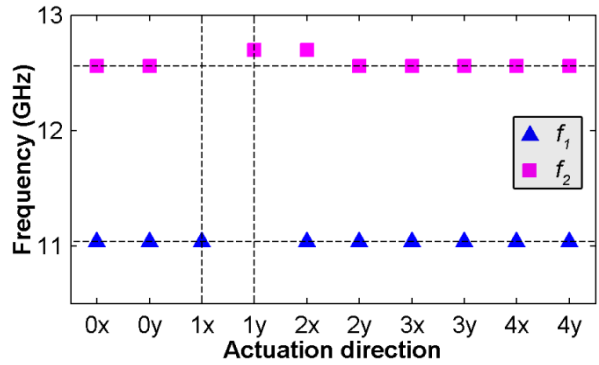

**Fig. 4.** Comparisons of the frequency values determined by FFT from the ten actuation directions shown in figure 3.

We now demonstrate that the decomposition of the initial single actuation along the two elementary orthogonal directions is identical to simultaneously exciting the NW along both of the elementary directions. To do so, we assume validity of the superposition principle such that a velocity actuation $v(z) = \lambda \sin(kz)$ in the $[\bar{1}1\bar{2}]$ direction as in Fig. 3 has the projections $v_1 = v\cos\theta$ and $v_2 = v\sin\theta$ along the $x$ and $y$-axes, respectively. Due to superposition, the vibration of the NW excited by $\vec{v}$ is equivalent to the vibration driven by $\vec{v}_1 + \vec{v}_2$, which is confirmed by numerical tests. As shown in Fig. 5a, the *EE* curve as well as the frequency spectrogram for a NW under two orthogonal actuations is very similar to the results in Figs. 2a and 2b. This demonstrates that the ratio of the amplitudes of the two vibration components in the frequency domain reflects the actual decomposition of the single actuation into two orthogonal directions. Considering the NW as a simple spring-mass system, the energy input due to the velocity actuation is $AE_i = Mv_i^2/2$, which means the ratio of the *EE* magnitude in the two elementary directions equals $AE_2/AE_1 = (v_2/v_1)^2 = \tan^2\theta$. Moreover, for an energy conserving system, the energy input should equal the maximum or the magnitude of *EE* that is derived from the FFT, meaning $AE_i = |Y_i(f_i)|$. Hence, we get $|\tan\theta| = \sqrt{|Y_2(f_2)|/|Y_1(f_1)|}$. The ideal value of $|\tan\theta|$ can easily be calculated according to the actuation angle $\theta$ according to Fig. 3, while the actual value is accessible from the frequency spectrogram of the MD simulations using the corresponding amplitude $|Y_i(f_i)|$. As shown in Fig. 5b, the values estimated from the FFT agree well with the theoretical values from Fig. 3.

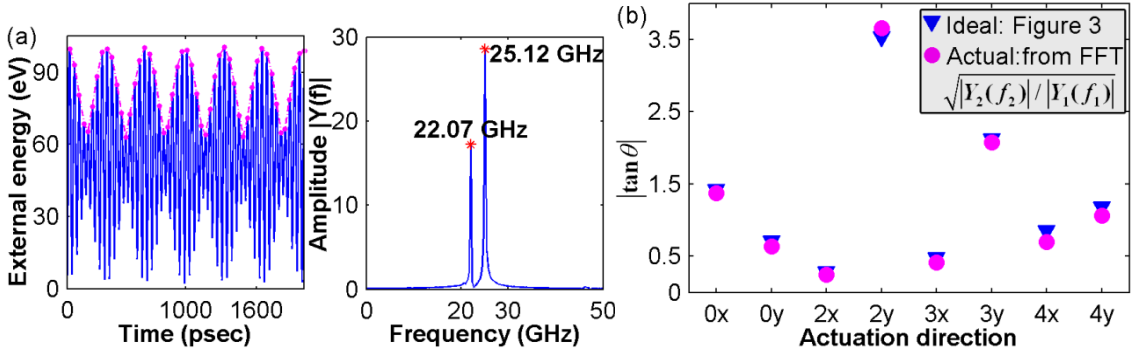

**Fig. 5.** (a) MD results from free vibration of [110] Ag NWs excited simultaneously by two velocity actuations applied along the two elementary directions. The left figure represents the time history of the external energy, while the right figure represents the frequency spectrum from FFT analysis. (b) The relations of the two velocity components in different cases. The ideal value is calculated from figure 3, with the actual value obtained from the FFT analysis of the MD simulation results.

We now address why [110] orientated FCC NWs exhibit this novel beat phenomenon. As is known, [110] orientated FCC NWs can be constructed by repeating two adjacent {110} atomic layers in the axial direction, where these two atomic layers possess an identical planar atomic arrangement. Therefore, a single (110) layer is taken as a fundamental repeating element. Considering the fact that the atomic system is discrete in nature, the (110) atomic layer is modeled as a plane composed of atomic hard spheres. Accordingly, the moment of inertia along the *x*-axis (as in Fig. 3) can be expressed as

$$I_x = \int_A y^2 dA = \sum \left( y_n^2 A_n + \pi r^4 / 4 \right) \quad (3)$$

where $A_n$ represents the projected area of the *n*th atom, $y_n$ is the *y* coordinate of atom *n*, and *r* is the atomic radius. Taking the C1 coordinate system as in Fig. 3, it is clear that due to the different atomic arrangements along the $[\bar{1}10]$ and [001] directions, $I_{[\bar{1}10]} \neq I_{[001]}$. However, since both of these directions are symmetry axes of the circular cross-section, the product moment of inertia equals zero, which means that the moments of inertia $I_{[\bar{1}10]}$ and $I_{[001]}$ are the two principal moments of inertia for the (110) atomic layer, which is the cross-sectional plane of the [110] Ag NW shown in Fig. 3. In other words, the beat vibration for [110] orientated FCC NWs is composed of vibrations in two orthogonal planes, in which the moments of inertia are maximum and minimum, respectively. This conclusion is consistent with that reported previously by Gil-Santos et al.[44], who found that when the cross-sectional symmetry of Si NWs is broken, a single planar vibration will split into two vibrations in orthogonal planes, which possess a maximum and minimum moment of inertia. Therefore, we conclude that the beat vibration originates from the asymmetric arrangement of atoms within the (110) atomic layer.

Another important question to resolve is the circumstances in which the beat phenomenon will disappear. If we consider a [110] NW with rectangular cross-section (this cross-sectional geometry was chosen to more easily quantify the surface effects) using the C1 coordination system, the NW will be enclosed by two {110} and two {001} surfaces. The effective flexural rigidity can be expressed as, while including the effects of surface elasticity[45]:

$$(EI)_x^* = EI_x + E_{(001)}bh^2/2 + E_{(\bar{1}10)}h^3/6 \tag{4}$$

where $b$ and $h$ are the cross-sectional lengths along the $x$ and $y$-axes, respectively, $E$ is the bulk Young's modulus of the [110] NW and $E_{(\bar{1}10)}$ and $E_{(001)}$ are the two surface Young's modulus, respectively. The moment of inertia can be derived from the discrete model presented in equation (3) as (see Supporting materials for derivation details):

$$\begin{cases} I_x = \lambda_1 + \lambda_2 M(2M+1)/3 - \lambda_2 M^2/2 \\ I_y = \lambda_1 + \lambda_2 N(2N+1)/6 - \lambda_2 N^2/4 \end{cases} \tag{5}$$

where $\lambda_1 = (M+1)(N+1)\pi r^4/4$, $\lambda_2 = a^2(M+1)(N+1)\pi r^2/2$, $a$ is the lattice constant, $r = a/(2\sqrt{2})$, $M$ and $N$ represent two integers corresponding to the column numbers of atoms along the $x$ and $y$-axes, i.e., $b = \sqrt{2}aN/2$, $h = Ma$. Assuming the vibrational properties of the NW follow the classical Euler-Bernoulli beam theory[41], the resonance frequency is:

$$f = \omega_n\sqrt{EI/\rho A}/(2\pi L^2) \tag{6}$$

where $\rho$ is the density, $A$ is the cross-sectional area of the NW and $\omega_n$ is the eigenvalue corresponding to different boundary conditions. According to equation (6), the ratio of the two frequency components can be expressed as $R_{xy} = f_y/f_x = \sqrt{(EI)_x^*/(EI)_y^*}$. Therefore, setting $h \approx b$ to model a square cross-section, we discuss the bounds to the beat phenomenon using the discrete model, where for the Ag NW, the surface Young's moduli are given as $E_{(\bar{1}10)} = -4.74$ N/m and $E_{(001)} = 1.22$ N/m, respectively[45, 46]. The bulk Young's modulus is taken as 76 GPa[47].

Figure 6a illustrates the predictions of $R_{xy}$ both including and neglecting surface effects in equation (4). For both cases, the values of $R_{xy}$ can be well described by a reciprocal function, which converges to one with increasing $b$. This fact indicates that the beat phenomenon will disappear when the cross-sectional size is too large to resolve the two frequencies, and corresponds to the size-dependent transition from a discrete to continuous system. While it is well-known that surface effects become stronger with decreasing NW cross-sectional size, it is seen in Fig. 6a that while the changing pattern is not impacted by surface effects, the value of $R_{xy}$ is obviously increased, and thus we can conclude that surface effects enhanced the beat phenomena. As compared in Fig. 6a, the value of $R_{xy} = f_y / f_x$ obtained from MD simulations (star markers) agree well with the predictions by the discrete model in equations. (4) and (5) (see Supporting materials for complete MD results). Hence, the decreasing $R_{xy}$ in Fig. 6a implies a decrease in the difference between the two frequency components. In other words, based on equation (6), the Young's moduli along the two orthogonal directions show a larger difference for NWs with smaller cross-sectional size, and thus the beat phenomenon is enhanced for smaller cross-section NWs. On the other hand, because the frequency is a reciprocal function of the length square, the frequency difference will decrease with an increase of the length or the aspect ratio. Figure 6b shows the absolute frequency difference $\Delta f = f_y - f_x$ as a function of the aspect ratio. As expected, $\Delta f$ decreases with an increase of the aspect ratio, which indicates the two distinct natural frequencies will become too close to distinguish for longer NW. In all, the beat phenomenon is expected to disappear when the NW's cross-sectional size, length or aspect ratio become large.

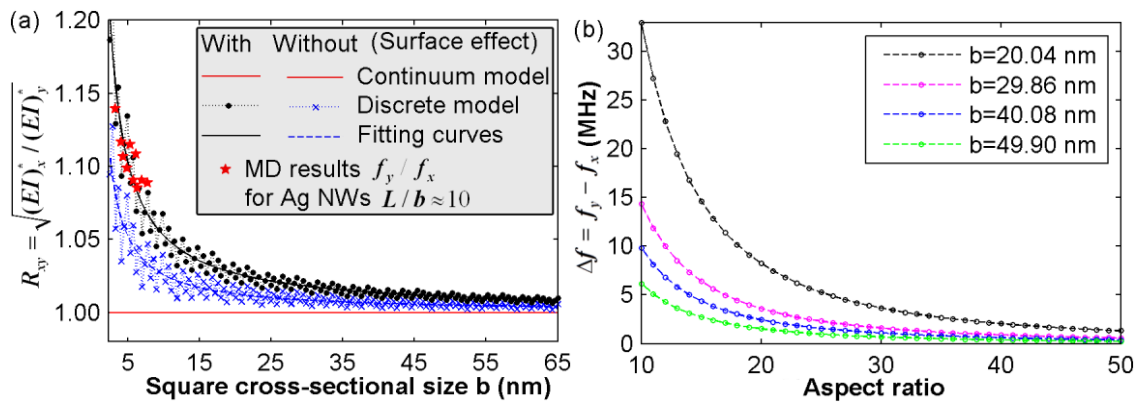

**Fig. 6.** (a) The change of $R_{xy}$ both includes and neglects surface effects for Ag NWs with a square cross-section. (b) The absolute frequency difference $\Delta f$ for the square Ag NW as a function of the aspect ratio.

We finally discuss the implications of the beat phenomena for interpreting the elastic properties, and specifically the Young's moduli, that are obtained from experimental resonance tests. We assume that the NWs in this work follow classical Euler-Bernoulli beam theory[41], i.e., equation (6). Substituting the two frequency values in Fig. 2b and using a classical flexural rigidity of $EI = E\pi D^4 / 64$ gives two different values for Young's modulus as 67.23 GPa and 87.09 GPa along the [$\bar{1}$10] and [001] directions, respectively ($\rho$ is chosen $1.05 \times 10^4$ kg/m$^3$ according to Shenoy[46]). The lower Young's modulus agrees well with the resonance experimental results from Cuenot et al.[48] as $67.5 \pm 2.1$ GPa, while the higher Young's modulus agrees well with the bulk values for Ag cited by previous researchers, e.g, 76 GPa by Jing et al.[47], 82.7 GPa by Chen et al.[49]. The following discussion assumes the existence of two close resonant frequencies for the NW, which correspond to two different Young's moduli.

In experimental studies of NW resonance, the actuation is often in the form of an AC electric field, where the resonance is detected by sweeping the actuation frequency until a maximum in the vibrational amplitude is observed. To model this situation analytically, we consider the NW to act as a damped mass-spring system along each elementary direction. If the AC electric field is modelled as a time harmonic force $F = F_0 \sin(\omega t)$, we can write its two projections onto the elementary directions as $F_x = F\cos\theta$ and $F_y = F\sin\theta$, where $\omega$ is the force frequency or sweep frequency, and $\theta$ is the actuation angle (see Fig. 3). Based on the Cauchy law, we have the ratio of the elastic constants of the springs $r = k_1/k_2 = E_1/E_2$ ($E_1 < E_2$). According to vibration theory [41], the amplitude of the stable vibration in each elementary direction can be expressed as $B_x = h_1 \cos\theta \big/ \sqrt{(1-\lambda_1^2)^2 + 4\zeta_1^2 \lambda_1^2}$ and $B_y = h_2 \sin\theta \big/ \sqrt{(1-\lambda_2^2)^2 + 4\zeta_2^2 \lambda_2^2}$, where $h_i = F_0/k_i$, $\zeta_i = n/\omega_{ni}$, $\lambda_i = \omega/\omega_{ni}$, $\omega_{ni}^2 = k_i/M$ and $2n = c/M$ with $c$ as the coefficient of damping force. Hence, according to linear superposition, the amplitude in the actuation direction can be obtained as $B_{xy} = B_x \cos\theta + B_y \sin\theta$. Let the relative amplitude $\beta_{xy} = B_{xy} k_2 / F_0$, hence (see Supporting materials for derivation details)

$$\beta_{xy} = \frac{\cos^2\theta}{\sqrt{(r-\lambda^2)^2 + 4\zeta^2 \lambda^2}} + \frac{\sin^2\theta}{\sqrt{(1-\lambda^2)^2 + 4\zeta^2 \lambda^2}} \qquad (7)$$

where $\lambda = \lambda_2$ and $\zeta = \zeta_2$. Taking the circular Ag NW (diameter of 6 nm and length of 34.71 nm) considered in the present work as an example, $r = 67.23/87.09 \approx 0.77$. Based on this value, and letting $\zeta = 0.01$ to model a slightly damped system, the amplitude-frequency relations for different actuation angles $\theta$ can be compared. Fig. 7a presents the variation of the relative amplitude with increasing relative sweep frequency. The two peak values represent the fundamental resonance frequencies along the two elementary orthogonal directions. As is seen, the difference between the two peak amplitudes decreases as the actuation angle $\theta$ approaches a certain value (around $\pi/4$). For instance, when $\theta = \pi/6$, the relative amplitudes at the two resonance frequencies equal 43.79 and 15.75, respectively. The significant difference in the relative amplitude would likely result in the resonance frequency with the larger amplitude being identified as the natural frequency of the NW. In other words, only one distinct frequency component would be detected for the actuation angle $\theta$ smaller than $\pi/6$ or larger than $\pi/3$, though the frequency corresponds to the smaller Young's modulus for $\theta < \pi/6$ and the larger Young's modulus for $\theta > \pi/3$ (see Supporting materials for more details).

Setting $dB_x/d\lambda_1 = 0$ and $dB_y/d\lambda_2 = 0$, the relative amplitudes at the two resonance frequencies can be obtained as $\beta_{r1} = B_{r1}k_2/F_0$ and $\beta_{r2} = B_{r2}k_2/F_0$ for the lower and higher frequency, respectively. Apparently, these two values are a function of both the actuation angle $\theta$ and the damping ratio $\zeta$. According to Fig. 7b, $\beta_{r1}$ (solid lines) decreases with an increase of the actuation angle while $\beta_{r2}$ (dashed lines) increases with an increase of the actuation angle. As shown in Fig. 7b, the critical actuation angle $\theta_{cr}$ when $\beta_{r1} = \beta_{r2}$ is always larger than $\pi/4$, which increases with an increase in the damping ratio. The critical actuation angle $\theta_{cr}$ also becomes larger than $\pi/4$ for smaller $r$ (see Supporting materials for more details). These facts indicate that the smaller Young's modulus will be measured experimentally if the resonance frequency with the highest amplitude is taken to be the natural frequency of the NW when the actuation angle $\theta$ is smaller than the critical value $\theta_{cr}$.

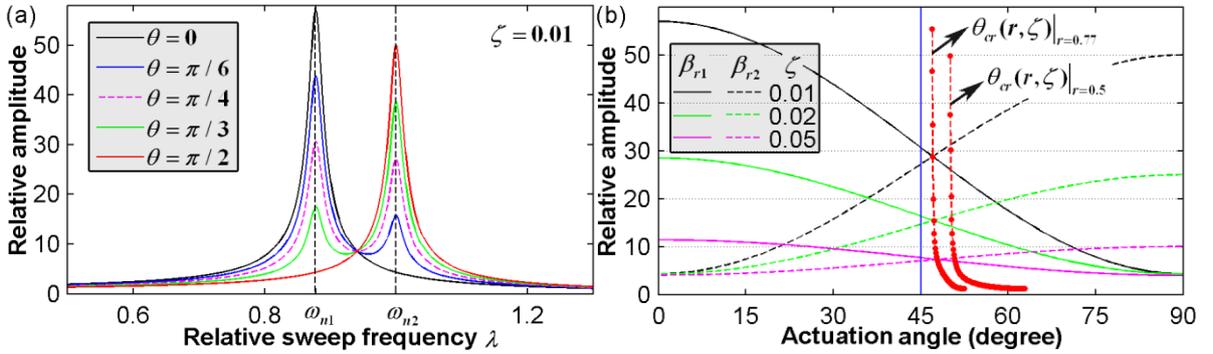

**Fig. 7.** (a) The relative oscillation amplitude as a function of the relative sweep frequency. (b) The relative oscillation amplitudes at the higher and lower resonance frequencies as a function of the actuation angle for damping ratios $\zeta = 0.01$, 0.02 and 0.05.

The above analysis and Fig. 7 suggests that for most experimental studies of NW resonance, one would observe only a single frequency component in spite of the existence of the beat phenomena, particularly when the damping ratio is very large or very small, and so long as the actuation angle does not equal about $\pi/4$. Furthermore, the lower frequency is more likely to be detected than the higher one due to its larger relative amplitude for actuation angles that are smaller than the critical actuation angle $\theta_{cr}$ (where $\theta_{cr}$ is dependent on the various factors discussed above, but it is always larger than $\pi/4$, as shown in Fig. 7b).

Before concluding, we note that there are some publications that report a similar beat phenomenon, for instance, Bai et al.[50] reported that the ZnO nanobelts exhibit dual-mode mechanical resonance, while Nam et al.[16] observed two normal vibration modes for GaN NWs. It is interesting that a diameter-dependent Young's modulus was reported by Nam et al. in Ref. 16, where a significant reduction in Young's modulus was reported for smaller diameter GaN NWs, which could also occur if the lower resonant frequency is consistently measured for smaller diameter NWs. We were unable to find other examples of beat phenomena reported in the literature, likely because most experiments search only for a single resonant frequency.

Most applications of NW-based NEMS currently assume that the NW oscillates in a single plane. However, researchers have exploited non-planar oscillations of NWs for new and intriguing application. For example, Conley et al.[51] recently showed that a NW resonator can suddenly transit from a planar motion to whirling, 'jump rope' like motion, which results from a non-planar vibration mode. Furthermore, Gil-Santos et al.[44] proposed a new approach for mass sensing and stiffness spectroscopy based on the non-planar vibration of Si NWs by breaking the symmetry of the NW. This approach was shown to enable the measurement of mass with zeptogram sensitivity, and discriminating variations in the Young's modulus of ~0.1 KPa per femtogram of sample. Our results suggest that due to the asymmetric cross-section, [110] orientated FCC metal NWs naturally exhibit a non-planar vibration, which implies they could be utilized for the non-planar sensing approach proposed by Gil-Santos et al.[44]. Therefore, it is expected that the beat phenomenon of metal nanowires could have significant application to NEMS that could operate in the non-planar regime.

**Conclusions**

In summary, we have utilized large scale MD simulations to report a novel beat phenomenon driven by a single actuation for [110] orientated Ag NWs. The beat phenomenon was found to be quite general, existing for different FCC materials (Cu, Au, Ni, Pd and Pt), temperatures, cross-sectional sizes, cross-sectional geometries, methods of inducing the NW oscillations, and boundary conditions. We developed a discrete moment of inertia model based on the hard sphere assumption to demonstrate that the beat phenomenon originates from the asymmetry in the lattice spacing in the (110) plane of the NW cross-section, which implies that it should exist for NWs of different axial orientations that have cross-sections with non-symmetric lattice spacings. Our results suggest that in many resonance experiments, a single frequency is expected to be detected in spite of the existence of the beat phenomena, particularly when the damping ratio is very large or very small. Furthermore, for a wide range of the actuation angle, the lower frequency is more likely to be detected, which implies that the elastic properties, and in particular the Young's modulus that is reported from experimental resonance tests of NWs may in fact be under predictions of the actual Young's modulus.

**Acknowledgement**

Support from the ARC Future Fellowship grant (FT100100172) and the High Performance Computer resources provided by the Queensland University of Technology are gratefully acknowledged.

**Supporting materials:** 1) Summary of all numerical testing results; 2) Some corresponding derivation details.